\documentclass[a4paper,10pt]{article}
\usepackage[latin1]{inputenc} % Encodage ISO-8859-1 des caractËres.
\usepackage[T1]{fontenc} % Package servant ? reconnaÓtre les accents de notre langue franÁaise pour les utilisateurs Linux.
\usepackage{graphics} % Pour inclure des figures en eps.
\usepackage{amssymb,amsmath} % Pour parler le langage des symboles mathÈmatiques et des Èquations compliquÈes...
\usepackage{times} % Pour utiliser des fonts de type times
\usepackage{vmargin} % C'est le package ? utiliser...
\usepackage[french]{babel}

\setmarginsrb{2.5cm}{2.5cm}{2.5cm}{2.5cm}{0cm}{0cm}{0cm}{0cm} % ...pour dÈfinir de nouvelle marges ? l'aide de cette commande...
\makeatletter
\renewcommand\section{\@startsection{section}{1}{\z@}% L? on s'attaque aux sections
{-2.5ex \@plus -1ex \@minus -.2ex}% l? on dÈfinit les espaces avant le titre de la section
{1.5ex \@plus .2ex}% l? c'est pour l'espace aprËs le titre de la section
{\reset@font\normalsize\bfseries}}% et enfin on s'amuse avec le style des titres et des numÈros de sections
\usepackage{caption} % Les commandes ci-aprËs sont contenues dans ce package.
\title{\textbf{Source laser intense pour le refroidissement du
$^{87}$Rb \\par doublement de fréquence d'un laser fibré télécom}}
\author{\normalsize P.-E. Pottie$^{\star}$, L. Longchambon, J.
Delaporte, R. Desbuquois, T. Liennard, V. Lorent, H. Perrin\\
\vspace{-0.16cm} \footnotesize{\textit{ Laboratoire de Physique
des Lasers, CNRS / Université Paris Nord-13,
Avenue J.-B. Clément, 93430 Villetaneuse
}}\\
\vspace{-0.16cm}
\footnotesize{\it{$\star$ pottie@univ-paris13.fr}}
}
\date{} % Ne retourne aucune date dans le document final.

\begin{document}
\maketitle
\begin{center}
\begin{minipage}[]{14cm}
\textbf{R\'{e}sum\'{e} }: Nous avons construit un système laser
pour le refroidissement du $^{87}$Rb bas\'{e} sur le doublement de
fr\'{e}quence en simple passage de sources laser fibr\'{e}es
t\'{e}l\'{e}com. Grâce aux d\'{e}veloppements technologiques
intensifs dont elles ont b\'{e}n\'{e}fici\'{e}, ces sources
puissantes possèdent d'excellentes caract\'{e}ristiques en terme
de stabilit\'{e} d'intensit\'{e} relative et de bande passante de
modulation. L'efficacit\'{e} accrue de doublement de fr\'{e}quence
des cristaux à retournement p\'{e}riodique nous a permis d'obtenir
jusqu'à 1.8W à 780 nm à partir de 10 W à 1.56 $\mu$m en simple
passage dans un cristal de ppLN:MgO. Cette technique peut être
\'{e}tendue au refroidissement du potassium (767
nm)\,\cite{Bourdel:2009} et à la r\'{e}alisation de pièges
dipolaires.

\textbf{Mots-cl\'{e}s} : Doublement de fr\'{e}quence, atomes
froids, sources laser.
\end{minipage}
\end{center}

\section{Introduction}
La technique de quasi-accord de phase par retournement
p\'{e}riodique des propri\'{e}t\'{e}s non-lin\'{e}aires dans
certains cristaux pr\'{e}sente l'avantage d'exploiter un
\'{e}lement unique, et le plus fort, du tenseur de
susceptibilit\'{e} di\'{e}lectrique du mat\'{e}riau. L'accord de
phase est r\'{e}alis\'{e} sans angle entre les ondes fondamentale (pompe) 
et harmonique,
\'{e}liminant le d\'{e}calage spatial des deux ondes, même sur de
grandes longueurs de propagation. Les rendements obtenus de
quelques $\%$/W avec ces types de cristaux sont bien
sup\'{e}rieurs aux rendements des cristaux monolithiques usuels. Ces forts
rendements ont permis d'atteindre des puissances de l'ordre du
watt en r\'{e}gime impulsionnel et
continu\,\cite{Juwiler:2003}-\cite{Vyatkin:2004}.

Pour un faisceau focalis\'{e} au centre d'un cristal
non lin\'{e}aire, en quasi-accord de phase, dans la limite des
faibles efficacit\'{e}s (pas de d\'{e}pl\'{e}tion de la pompe), et
en n\'{e}gligeant les effets photor\'{e}fractifs et de lentille
thermique, l'efficacit\'{e} de doublement de fr\'{e}quence,
définie comme le rapport de la puissance doublée $P_{2
\omega}$ au carré de la puissance de la pompe $P_{\omega}$,
s'\'{e}crit\,:
\begin{equation}\label{eq:p2omega}
\frac{P_{2\omega}}{P_{\omega}^{2}}= \eta l_{c}
d_{\mathrm{eff}}^{2} h\exp(-\alpha_{2\omega}l_{c})
\end{equation}
où $\eta$ est un facteur de qualit\'{e} d\'{e}pendant de la
longueur d'onde fondamentale et des indices extraordinaires du
cristal, $l_{c}$ la longueur du cristal , et $d_{\mathrm{eff}}$
est le coefficient effectif de doublement de fr\'{e}quence. $h$
est la fonction de focalisation de
Boyd-Kleinman\,\cite{Kleinman:1968}. $\alpha_{2\omega}$ traduit
les pertes lin\'{e}aires à la fr\'{e}quence harmonique.

\section{Montage exp\'{e}rimental}
Nous avons choisi un cristal de niobate de lithium dop\'{e} MgO à 5$\%$
produit par HCP Photonics, pour son fort coefficient non
lin\'{e}aire de 17.6pm/V\cite{Myers:1995}. Le cristal mesure 50$\times$3$\times$0.5
mm. Il est maintenu dans les mâchoires d'un four à une
temp\'{e}rature voisine de 89\degre C à 0.1\degre~C près pour
assurer le quasi-accord de phase aux deux longueurs d'onde.

Le laser pompe est une diode fibr\'{e}e DFB FITEL FOL15DCW,
\'{e}mettant plus de 1~mW à 1560~nm, saturant l'amplificateur en entrée. 
Le laser injecte un
amplificateur fibr\'{e} monomode à maintien de polarisation de 40~dB 
de gain (Kheopsys), saturant en sortie à 10~W. Le laser
amplifi\'{e} est collimat\'{e} en sortie par une lentille de
focale 8~mm et trait\'{e}e anti-reflet (A/R)) à 1560~nm, connectoris\'{e}e à la
fibre cliv\'{e}e droite de sortie de l'amplificateur. Le col de
725 microns est localis\'{e} 10~cm en arrière de la sortie de
fibre. Le faisceau est ensuite focalis\'{e} au centre du cristal
par une lentille de focale 100~mm, plac\'{e}e sur une platine de
translation microm\'{e}trique. On obtient un col de 75~microns
plac\'{e} au centre du cristal, proche de l'optimum calcul\'{e}
par la fonction de Boyd-Kleinman\,\cite{Kleinman:1968}. Un miroir
et une platine de micro-positionnement 5 axes permettent
l'alignement de la pompe et de l'axe lent du cristal sur toute sa
longueur, ainsi qu'un bon positionnement du col au centre du
cristal. Une lentille trait\'{e}e A/R à 780~nm fait l'image à
l'infini de ce col. Le faisceau est diffracté à l'ordre 1 par un modulateur acousto-optique
puis est inject\'{e} dans une fibre monomode à maintien de polarisation
avec 90$\%$ d'efficacité. 

\section{Résultats expérimentaux}
\begin{figure}[t]
\begin{center}
\resizebox{0.45\columnwidth}{!}{\includegraphics{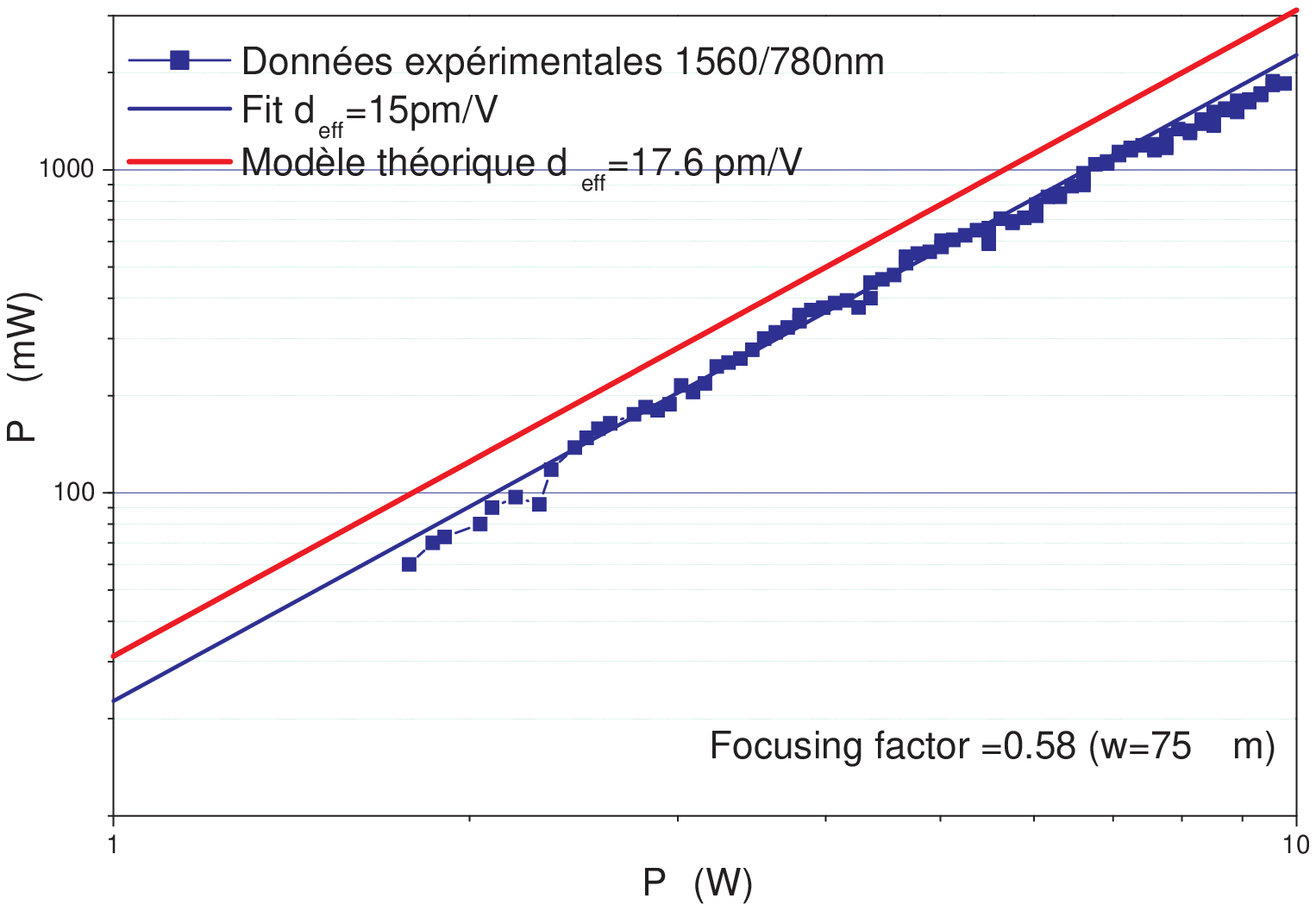}}
\hspace*{0.2cm}
\resizebox{0.45\columnwidth}{!}{\includegraphics{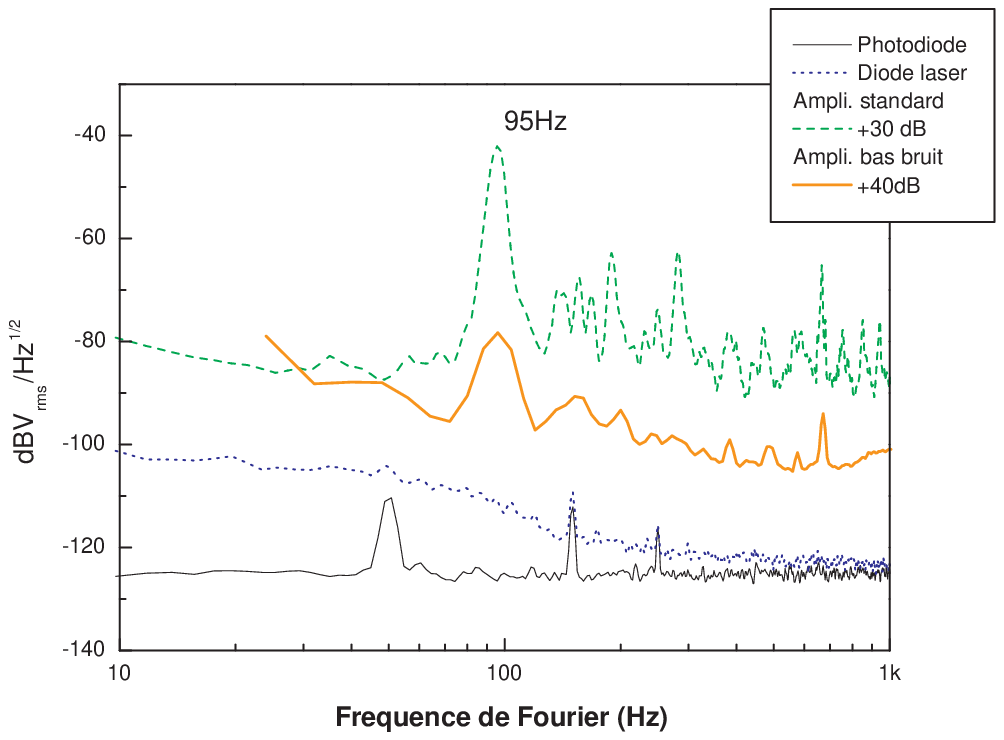}}
\vspace*{-5mm}\caption{(gauche) Puissance de sortie en fonction de la puissance de 
pompe. (droite) Spectre de bruit d'intensité du laser à 1560~nm.}
\label{resultats}
\end{center}
\end{figure}

Nous avons obtenu à la sortie du laser doublé jusqu'à 1.8~W en continu à 10~W de puissance de
pompe ({Fig.~\ref{resultats}} à gauche). En ajustant la courbe expérimentale par la formule
(\ref{eq:p2omega}), on obtient un coefficient non linéaire effectif
$d_{\mathrm{eff}}$=15~pm/V, lég\`{e}rement en dessous de la valeur
théorique. On a estimé la largeur en fréquence du laser pompe en
faisant le battement entre deux diodes lasers semblables à 1560~nm. 
La largeur du battement est de 1.7~MHz. On en déduit la
largeur en fréquence du laser à 780~nm, égale à 2.4~MHz. A l'aide d'un montage d'absorption saturée, 
nous avons asservi le laser doublé en fréquence sur une transition hyperfine de la raie 
D2 du rubidium, en rétro-agissant sur le courant de la diode à 1560~nm. Nous 
avons obtenu un piège magnéto-optique de $10^{9}$ atomes, limité par la taille
du col des faisceaux de refroidissement et par la pression de vapeur de rubidium dans l'enceinte. 
Cette source servira
également à la réalisation d'un MOT-2D+\cite{Conroy:2003}.

Dans la perspective d'utiliser ce type de source pour la réalisation d'un piége
dipolaire (pour le rubidium ou le potassium par exemple), 
nous avons caractérisé le bruit d'intensité du syst\`{e}me
laser par analyse de Fourier des fluctuations d'intensité aux
bornes d'une photodiode en montage transimpédance, pour deux
versions de l'amplificateur fibré. Nous avons mis en
évidence l'existence d'un pic de bruit à 95~Hz et ses harmoniques
impaires sur la version standard du fabricant, dû aux vibrations
acoustiques des ventilateurs utilisés dans le chassis. Grâce à une
collaboration avec le fabricant, une version
bas bruit a été mise au point, réduisant fortement ce pic de bruit 
(Fig.~\ref{resultats}, à droite). La variance des
fluctuations relatives d'intensité à 10~W est de - 72~dB sur la
bande 1 Hz-102~kHz. La densité spectrale de bruit relatif d'intensité est inférieure à 
$-110$~dB/Hz à 10~W dans la bande 1-10~kHz. Pour un piège de fréquence d'oscillation 
5~kHz, le taux de chauffage
paramétrique correspondant est inférieur à 2.4$\times$10$^{-3}$/s
\cite{Savard:1997}. Ces valeurs très faibles indiquent que le bruit d'intensité n'est pas
limitant pour la réalisation de pièges dipolaires avec ce type de sources laser.

\section{Remerciements}
\noindent Le laboratoire de physique des Lasers est une unité
mixte CNRS/Université Paris Nord (UMR 8538). Les auteurs remercient l'IFRAF et le 
PPF "Manipulation d'atomes froids par des lasers de puissance"
pour leur soutien financier, et Olivier Lopez pour son aide
précieuse et le prêt de matériel.

\end{document}